\documentstyle[aps]{revtex}
\begin{document}
\draft
\def\lsim{\lower.5ex\hbox{$\; \buildrel < \over \sim \;$}}
\def\gsim{\lower.5ex\hbox{$\; \buildrel > \over \sim \;$}}
\title{Solution of Dirac equation around a spinning Black Hole}
\author{Banibrata Mukhopadhyay\\ 
       and\\
  Sandip K. Chakrabarti\\}

\address{Theoretical Astrophysics Group\\
     S. N. Bose National Centre For Basic Sciences,\\
     JD Block, Salt Lake, Sector-III, Calcutta-700091\\}        
\thanks{e-mail: bm@boson.bose.res.in  \& chakraba@boson.bose.res.in\\}
\maketitle 

\setcounter{page}{1}
\noindent{ Nuclear Physics B: In Press}

\baselineskip = 24 true pt
	
\def\ch{\lower-0.55ex\hbox{--}\kern-0.55em{\lower0.15ex\hbox{$h$}}}
\def\lh{\lower-0.55ex\hbox{--}\kern-0.55em{\lower0.15ex\hbox{$\lambda$}}}

\begin{abstract}

Chandrasekhar separated the Dirac equation for spinning and massive particles 
in Kerr geometry in radial and angular parts. Chakrabarti 
solved the angular equation and found the corresponding 
eigenvalues for different Kerr parameters. The radial 
equations were solved asymptotically by Chandrasekhar. 
In the present paper, we use the WKB approximation
to solve the spatially complete radial equation and 
calculate analytical expressions of radial wave functions for a set of Kerr and 
wave parameters. From these solutions we obtain local values of
reflection and transmission coefficients.

\end{abstract}
\vskip0.5cm
\vskip0.5cm


\section*{ I. INTRODUCTION}

One of the most important solutions of Einstein's equation is that
of the spacetime around and inside an isolated black hole.
The spacetime at a large distance is flat and Minkowskian
where usual quantum mechanics is applicable, while the spacetime
closer to the singularity is so curved that no satisfactory
quantum field theory could be developed as yet. An intermediate
situation arises when a weak perturbation (due to, say, gravitational,
electromagnetic or Dirac waves) originating from
infinity impinges on a black hole, interacting with it.
The resulting wave is partially transmitted into the
black hole through the horizon and partially scatters off
to infinity. In the linearized (`test field') approximation
this problem has been attacked in the past by several authors [1-4].
The master equations of Teukolsky [2] which govern these linear 
perturbations for integral spin (e.g., gravitational and electromagnetic) 
fields were solved numerically by Press \& Teukolsky [5] and 
Teukolsky \& Press [6]. While the equations governing the massive 
Dirac particles were separated by Chandrasekhar [3]. So far, only the 
angular eigenfunction and eigenvalue (which happens to be the separation constant)
have been obtained [7]. Particularly interesting is the fact that 
whereas gravitational and electromagnetic radiations were found 
to be amplified in some range of incoming frequencies,
Chandrasekhar [4] predicted that no such amplifications should
take place for Dirac waves because of the very nature of the
potential experienced by the incoming fields. However,
these later conclusions were drawn using asymptotic
solutions and no attempt has so far been made to determine the
nature of the radial wave functions, both incoming and outgoing,
for the Dirac wave perturbations. He also speculated that one needs
to look into the problem for negative eigenvalues ($\lh$) where
one {\it might} come across super-radiance for Dirac waves.

In the present paper, we revisit this important problem to study the
nature of the radial wave functions as a function of the
Kerr parameter, rest mass and frequency of incoming particle.
We also verify that super-radiance is indeed absent for the Dirac field.
Unlike the works of Press \& Teukolsky [5] and Teukolsky \& Press [6]
where numerical (shooting) methods were used to solve the master equations governing
gravitational and electromagnetic waves, we use an approximate analytical method
for the massive Dirac wave. The details of the method would be presented below.

The plan of the paper is as follows: in the next Section, we present
the equation governing the Dirac waves (waves for
half-integral massive spin particles) as they were separated into radial and angular
co-ordinates. We then briefly present the nature of the angular eigenvalues and
eigenfunctions. In \S 3, we present our method of solution 
and present the {\it spatially complete} radial wave functions. Finally, 
in \S 4, we draw our conclusions.

\section*{II. THE DIRAC EQUATION IN KERR GEOMETRY}

Chandrasekhar [3] separated the Dirac equation in Kerr geometry into
radial ($R$) and angular ($S$) wave functions. Below, we present these equations
from Chandrasekhar [4] using the same choice of units: we choose
$\ch=1=G=c$.

The equations governing the radial wave-functions $R_{\pm \frac{1}{2}}$
corresponding to spin $\pm \frac{1}{2}$ respectively are given by:

$$
\Delta^{\frac{1}{2}}{\cal D}_{0} R_{- \frac{1}{2}}
= ( \lh + i m_p r) \Delta^{\frac{1}{2}} {R}_{+ \frac{1}{2}} ,
\eqno{(1a)}
$$

$$
\Delta^{\frac{1}{2}} {\cal D}_{0}^{\dag} \Delta^{1 \over 2}
{R}_{+{\frac{1}{2}}} = ( \lh - i m_p  r)  {R}_{-{1 \over 2}} ,
\eqno{(1b)}
$$

where, the operators ${\cal D}_n$ and ${\cal D}_{n}^{\dag}$ are given by,

$$
{\cal D}_n = \partial_{r} + \frac {i K} {\Delta} + 2n \frac{(r-M)} {\Delta} ,
\eqno{(2a)}
$$

$$
{\cal D}_n^{\dag} = \partial_{r} - \frac {i K} {\Delta} + 2n \frac{(r-M)} {\Delta} ,
\eqno{(2b)}
$$
and
$$
\Delta = r^2 + a^2 - 2Mr ,
\eqno{(3a)}
$$
$$
K=(r^2 + a^2)\sigma + am .
\eqno{(3b)}
$$
Here, $a$ is the Kerr parameter, $n$ is an integer or half integer, $\sigma$ is the
frequency of incident wave, $M$ is the mass of the black hole, $m_p$ is the rest mass of
the Dirac particle, $\lh$ is the eigenvalue of the Dirac equation and $m$ is the azimuthal quantum number.

The equations governing the angular wave-functions $S_{\pm \frac{1}{2}}$ corresponding
to spin $\pm \frac{1}{2}$ respectively are given by:

$$
{\cal L}_{1 \over 2} S_{+ {1 \over 2}} = - (\lh -a m_p \cos \theta) S_{- {1 \over 2}}
\eqno{(4a)}
$$

$$
{\cal L}_{1 \over 2}^{\dag} S_{- {1 \over 2}} = + (\lh +a m_p \cos \theta) S_{+ {1 \over 2}}
\eqno{(4b)}
$$

where, the operators ${\cal L}_n$ and ${\cal L}_{n}^{\dag}$ are given by,

$$
{\cal L}_n = \partial_\theta + Q + n \cot \theta ,
\eqno{(5a)}
$$

$$
{\cal L}_n^{\dag} = \partial_\theta - Q + n \cot \theta
\eqno{(5b)}
$$
and
$$
Q=a \sigma \sin \theta + m \  {\rm cosec} \  \theta .
\eqno{(6)}
$$

Note that both the radial and the angular sets of equations i.e., eqs. 1(a-b)  and eqs. 4(a-b)
are coupled equations. Combining eqs. 4(a-b), one obtains the angular eigenvalue equations for the
spin-$\frac{1}{2}$ particles as [7]
$$
\left [{\cal L}_{1 \over 2} {\cal L}_{1 \over 2}^{\dag} +
\frac{a m_p \sin \theta}{\lh + a m_p \cos \theta} {\cal L}_{1\over 2}^{\dag} +
(\lh^2 - a^2 m_p^2 \cos^2 \theta) \right ] S_{-{1 \over 2}} = 0 .
\eqno{(7)}
$$

There are exact solutions of this equation for the eigenvalues $\lh$ and the eigenfunctions
$S_{-{1\over 2}}$ when $\rho=\frac{m_p}{\sigma}=1$ in terms of the orbital quantum number $l$
and azimuthal quantum number $m$. These solutions are [7]:
$$
\lh^2 = (l+\frac{1}{2})^2 + a \sigma ( p+ 2m) + a^2 \sigma^2 \left [1-\frac{y^2}{2(l+1)
+a\sigma x} \right ] ,
\eqno{(8)}
$$
and
$$
{}_{1\over 2}S_{lm} =
{}_{1\over 2}Y_{lm} - \frac{a\sigma y}{2(l+1)+a\sigma x} {}_{1\over 2}Y_{l+1 m}
\eqno{(9)}
$$
where,
$$
p=F(l,l); \ \ \ x=F(l+1,l+1); \ \ \ y=F(l,l+1)
$$
and
$$
F(l_1,l_2)=[(2l_2+1)(2l_1+1)]^{\frac{1}{2}} <l_2 1 m 0|l_1 m>
$$
$$
[<l_2 1 \frac{1}{2} 0|l_1 \frac{1}{2}> +(-1)^{l_2-l}<l_2 1 m 0|l_1 m>
[<l_2 1 \frac{1}{2} 0|l_1 \frac{1}{2}> +(-1)^{l_2-l}
\rho \sqrt{2} <l_2 1 -\frac{1}{2} 1|l_1 \frac{1}{2}>]] .
\eqno{(10)}
$$
with $<....|..>$ are the usual Clebsh-Gordon coefficients.
For other values of $\rho$ one has to use perturbation theories. Solutions upto sixth order
using perturbation parameter $a\sigma$ is given in Chakrabarti [7]. The eigenfunctions
$\lh$ are required to solve the radial equations which we do now.

This radial equations 1(a-b) are in coupled form. One can decouple them and express
the equation either in terms of spin up or spin down wave functions $R_{\pm \frac{1}{2}}$
but the expression loses its transparency. It is thus advisable to use the approach of
Chandrasekhar [4] by changing the basis and independent variable $r$ to
$$
r_{*} = r + \frac{2M r_+ + am/\sigma} {r_+ - r_-} {\rm log}
\left({r \over r_+} - 1\right) - \frac{2M r_- + am/\sigma} {r_+ - r_-} {\rm log}
\left({r \over r_{-}} - 1\right),
\eqno{(11)}
$$
(for $r > r_{+}$),
$$
{d \over dr_{*}} = {\Delta \over \omega^{2}} {d \over dr},
\eqno{(12)}
$$
$$
\omega^2 = r^2 + \alpha^2; \ \ \ \ \ \ \alpha^2 = a^2 + am/\sigma,
\eqno{(13)}
$$
to transform the set of coupled equations 1(a-b)
into two independent one dimensional wave equations given by:
$$
\left({d \over dr_{*}} -  i \sigma\right)P_{+ {1 \over 2}} = \frac {\Delta^{1 \over 2}}
{\omega^2} (\lh - i m_p r) P_{- {1 \over 2}}
\eqno{(14)}
$$
$$
\left({d \over dr_{*}} + i \sigma\right)P_{- {1 \over 2}} = \frac{\Delta^{1 \over 2} } {\omega^2}
(\lh + i m_p r) P_{+ {1 \over 2}} .
\eqno{(15)}
$$
Here, ${\cal D}_{0} = {\omega^{2} \over \Delta} ({d \over dr_*} + i \sigma)$
and ${\cal D}^{\dagger}_{0} ={\omega^2 \over \Delta} ({d \over dr_{*}} - i\sigma)$
were used and wave functions were redefined as $R_{- {1 \over 2}} = P_{- {1 \over 2}}$ and
$\Delta^{1 \over 2} R_{+ {1 \over 2}} = P_{+ {1 \over 2}}$.

We now define a new variable,
$$
\theta = tan^{-1} (m_p r/{\lh})
\eqno{(16)}
$$
which yields,
$$
\cos \theta = \frac{\lh} {\surd(\lh^2 + m_p^2 r^2)}, \ \ {\rm and}
\ \ \ \sin \theta = \frac{m_p r} {\surd(\lh^2 + m_p^2r^2)}
$$
and
$$
(\lh \pm i m_p r) = exp ({\pm i \theta}) \surd ({\lh}^{2} + m_p^2 r^2),
\eqno{(17)}
$$
so the coupled equations take the form,
$$
\left({d \over dr_{*}} - i \sigma \right) P_{+ {1 \over 2}} = \frac{\Delta^{1 \over 2}}
{\omega^2}({\lh}^2 + m_p^{2} r^2)^{1/2}
P_{- {1 \over 2}} exp\left[-i \tan^{-1} \left({{m_p r} \over {\lh}}\right)\right],
\eqno{(18a)}
$$
and
$$
\left({d \over dr_{*}} + i \sigma\right)P_{- {1 \over 2}} = \frac{\Delta^{1 \over 2}}
{\omega^2}({\lh}^{2} + m_p^{2} r^2)^{1/2} P_{+ {1 \over 2}} exp\left[ i \tan^{-1} \left({{m_p r} 
\over {\lh}}\right)\right].
\eqno{(18b)}
$$

Then defining,
$$
P_{+ {1 \over 2}} = \psi_{+ {1 \over 2}}\  exp\left[-{1 \over 2} i\  tan^{-1} \left({{m_p r} 
\over \lh}\right)\right]
\eqno{(19a)}
$$
and
$$
P_{- {1 \over 2}} = \psi_{- {1 \over 2}}\  exp\left[+{1 \over 2} i \ tan^{-1} \left({{m_p r} 
\over \lh}\right)\right],
\eqno{(19b)}
$$
we obtain,
$$
{{d{\psi}_{+ {1 \over 2}}} \over dr_{*}} - i \sigma \left(1 + {\Delta \over \omega^2}{{\lh m_p} \over 2\sigma}
{1 \over {\lh^2 + m_p^2 r^2}}\right){\psi}_{+ {1 \over 2}} = 
{\Delta^{1 \over 2} \over \omega^2}(\lh^2 + m_p^2 r^2)^{1/2}
\psi_{-{1 \over 2}}
\eqno{(20a)}
$$
and
$$
{{d{\psi}_{- {1 \over 2}}} \over dr_{*}} + i \sigma \left(1 + {\Delta \over \omega^2}{{\lh m_p} \over 2\sigma}
{1 \over {\lh^2 + m_p^2 r^2}}\right){\psi}_{- {1 \over 2}} =
{\Delta^{1 \over 2} \over \omega^2}(\lh^2 + m_p^2 r^2)^{1/2} \psi_{+{1 \over 2}}.
\eqno{(20b)}
$$

Further choosing $\hat{r}_* = r_{*} + {1 \over 2\sigma} {\rm tan}^{-1}({m_p r \over \lh})$ so that
$d{\hat r}_* = (1 + {\Delta \over \omega^2} {{\lh m_p} \over 2 \sigma} {1 \over {\lh^2 + m_p^2 r^2}})dr_*$,
the above equations become,
$$
\left(\frac{d} {d{\hat r}_*} - i \sigma\right){{\psi}_{+ {1 \over 2}}} = W {{\psi}_{- {1 \over 2}}},
\eqno{(21a)}
$$
and
$$
\left(\frac {d} {d{\hat r}_*} + i \sigma\right){{\psi}_{- {1 \over 2}}} = W {{\psi}_{+ {1 \over 2}}}.
\eqno{(21b)}
$$
where,
$$
W = \frac{\Delta^{1 \over 2} (\lh^{2} + m_p^2 r^2)^{3/2} } { \omega^2 (\lh^2 + m_p^2 r^2)
+ \lh m_p \Delta/2\sigma}.
\eqno{(22)}
$$

Now letting $Z_{\pm} = \psi_{+ {1 \over 2}} \pm \psi_{-{1 \over 2}}$
we can combine the differential equations to give,

$$
\left(\frac{d} {d{\hat r}_*} - W\right) Z_+ = i \sigma Z_- ,
\eqno{(23a)}
$$

and

$$
\left(\frac{d} {d{\hat r}_*} + W\right) Z_- = i \sigma Z_+ .
\eqno{(23b)}
$$

>From these equations, we readily obtain a pair of independent one-dimensional wave equations,

$$
\left(\frac{d^2} {{d {\hat r}_*}^2} + \sigma^2\right) Z_\pm = V_\pm Z_\pm .
\eqno{(24)}
$$
where, $V_{\pm} = W^{2} \pm {dW \over d\hat{r}_{*}}$
$$
={{\Delta^{1 \over 2}(\lh^{2} + m_p^{2} r^{2})^{3/2}} \over {[ \omega^{2}(\lh^{2} + m_p^{2}
r^{2}) + \lh m_p \Delta/2 \sigma]^{2}}}[\Delta^{1 \over 2}(\lh^{2} + m_p^{2} r^{2})^{3/2} \pm
 ((r-M)(\lh^{2} + m_p^{2} r^{2}) + 3m_p^{2} r \Delta)]
$$

$$
\mp {{\Delta^{3 \over 2}(\lh^{2} + m_p^{2} r^{2})^{5/2}} \over {[ \omega^{2}(\lh^{2} + m_p^{2}
r^{2}) + \lh m_p \Delta/2 \sigma]^{3}}}[2r(\lh^{2} + m_p^{2} r^{2}) + 2 m_p^{2} \omega^{2} r +
 \lh m_p (r-M)/\sigma] .
\eqno{(25)}
$$

One important point to note: the transformation of spatial co-ordinate $r$ to
$r_{*}$ (and ${\hat{r}}_{*}$) is taken not only for mathematical simplicity
but also for a physical significance. 
When $r$ is chosen as the radial co-ordinate, the decoupled equations
for independent waves show diverging behaviour. 
However, by transforming those in terms of $r_{*}$ (and ${\hat{r}}_{*}$) 
we obtain well
behaved functions. The horizon is shifted from $r=r_+$ to ${\hat {r}}_{*}= -\infty$ unless
$\sigma \leq \sigma_s=-am/2Mr_+$ (eq. 11). In this connection, it is customary to define
$\sigma_c$ where $\alpha^2=0$ (eq. 13). Thus, $\sigma_c=-m/a$. If $\sigma \leq \sigma_s$,
the region is expected to be super-radiant [4] because for integral
spin particles for $\sigma \leq \ sigma_s$ there exhibit super-radiation.

\section*{III. SOLUTION OF THE RADIAL EQUATION}

Out of the total physical parameter space, in one region (region I)
the total energy of the particle is always greater than the height of 
potential barrier and in the other region (region II) the energy 
is less than of the maximum height of the potential barrier. 
In region II, the wave hits the wall of barrier and tunnels through it. 
One has to treat these two cases a little differently.

The usual WKB approximation [8] is used to obtain the zeroth order
solution. We improve the solution by properly incorporating the inner
and outer boundary conditions. 
After establishing the general solution, we present here the solution of eq. 
(4) for three sets of parameters as illustrative examples.
For those examples the choice of parameters is 
made in such a way that there is 
a significant interaction between the particle and 
the black hole, i.e., when the Compton
wavelength of the incoming wave is of the same order as the radius of
the outer horizon of the Kerr black hole. So,
$$
\frac{G [M + \sqrt(M^2 - a^2)]}{c^2} \sim \frac{\ch}{{m_p}c}.
\eqno{(26)}
$$
We choose as before $G=\ch=c=1$, so
$$
m_p \sim \frac{1}{[M + \sqrt(M^2 - a^2)]}.
\eqno{(27)}
$$
Similarly, the frequency of the incoming
particle (or wave) should be of the same order as the inverse 
of the light crossing time of the radius of the black hole, i.e., 
$$
\frac{c^3}{G [M + \sqrt(M^2 - a^2)]} \sim \sigma .
\eqno{(28)}
$$
Using the same units as before, we can write,
$$
m_p \sim \sigma \sim [M + \sqrt(M^2 - a^2)]^{-1}.
\eqno{(29)}
$$
In principle, however, one can choose any values of $\sigma$ and $m_p$
for a particular black hole and the corresponding solution is possible. 

One can easily check from equation (25) that for $r \rightarrow \infty$
(i.e., ${\hat{r}}_* \rightarrow \infty$) $V_\pm \rightarrow m_p^2$. 
So we expand the total parameter space in terms of the
frequency of the particle (or wave), $\sigma$ and the rest mass
of the particle, $m_p$. It is clear that in half of the parameter
space spanned by $\sigma-m_p$ where, $\sigma<m_p$, 
particles are released at finite
distance with so little energy that they cannot escape to infinity.
In this case,  the total energy $\sim \sigma^2$ of the incoming particle at a large distance
is less than the potential energy of the system.
We will not discuss solutions in this region.
The rest of the parameter space ($\sigma\ge m_p$) is divided into two
regions -- I: $E>V_{m}$ and II: $E<V_{m}$, where $E$ is the total energy
of the incoming particle and $V_{m}$ is the maximum of the
potential. In Region I, the wave is
{\it locally} sinusoidal because the wave number $k$ is real
for the entire range of ${\hat r}_*$. In Region II, on the other hand,
the wave is decaying in some region when $E <V$, i.e., where
the wave `hits' the potential barrier and in the rest of the
region, the wave is propagating. We shall show
solutions in these two regions separately. In
Region-I whatever be the physical parameters, the energy of the particle
is always greater than the potential energy and the WKB approximation is
generally valid in the whole range (i.e. $\frac{1}{k^2}\frac{dk}
{d\hat{r}_*} << 1$). In cases of Region-II, the energy of the particle
is always less than the maximum height of potential barrier. Thus, at 
two points (where, $k=0$) the total energy matches the potential energy and
in the neighbourhood of those two points the WKB approximate method is not valid.
They have to be dealt with separately. In Fig. 1, we show
contours of constant  $w_{max}=$max($\frac{1}{k^2}\frac{dk}{d\hat{r}_*}$)
for a given set  ($\sigma, m_p$) of parameters. The labels
show the actual values of $w_{max}$. Clearly, in most of the 
parameter regions the WKB approximation is safely 
valid for any value of $\hat{r}_*$. One has to
employ a different method (such as using Airy Functions, see below)
to find solutions in those regions where $w_{max}$ attains a large value
which indicates the non-validity of WKB method.

\subsection*{ Solutions of Region I} 

We re-write equation (24) as,
$$
{d^{2}Z_{+} \over d{\hat{r}}_{*}^{2}} + (\sigma^{2} - V_{+}) Z_{+} = 0 .
\eqno{(30)}
$$

This is nothing but the Schr\"odinger equation with total energy of the wave
$\sigma^2$. This can be solved by regular WKB method.

Let $k ({\hat r}_*) = \surd(\sigma^{2} - V_{+})$,
$u ({\hat r}_*) = \int k({\hat r}_*) d {\hat r}_* + {\rm constant}$. 
$k$ is the wavenumber of the incoming wave and $u$ as the {\it Eikonal}.
The solution of the equation (30) is,
$$
Z_{+} = \frac{A_{+}} {\surd k} exp (i u) + \frac {A_{-}} {\surd k} exp (- i u) .
\eqno{(31)}
$$
with 
$$
A_+^2+A_-^2 = k.
\eqno{(32)}
$$ 
The motivation of equation (32) is to impose the WKB method at the
each space point so that sum of the transmission and reflection 
coefficients are same at each location.
In this case  $\sigma^{2} > V_{+}$ all along and also ${1 \over k} 
{dk \over d{\hat{r}}_{*}} << k$, so the WKB approximation 
is valid in the whole region. 

It is clear that a standard WKB solution where  $A_+$ and $A_-$ are kept 
constant throughout,
can not be accurate in whole range of ${\hat{r}}_*$, 
since the physical inner boundary condition on the horizon
must be that the reflected component is negligible there (since there 
the potential barrier height goes down to zero).
Thus the WKB approximation requires a slight modification in which a
spatial dependence of $A_\pm$ is allowed. On the other hand,
at a large distance, where the WKB is strictly valid, $A_+$ and $A_-$
should tend to be constants, and hence their difference is also a constant:
$$
A_+-A_-=c.
\eqno{(33)}
$$
Here, one can choose also the sum of $A_+$ and $A_-$ are constant 
instead of difference as equation (33), but the final result will not
be affected. Here, $c$ is determined from the WKB solution at a
large distance. For simplicity we choose $A_\pm$s 
are real. This along with (32) gives,
$$
A_\pm (r) = \pm {c \over 2} + {\sqrt{[2k(r) - c^{2}]} \over 2} .
\eqno{(34)}
$$
This spatial variation, strictly valid at large distances only,
should not be extendible to the horizon without
correcting for the inner boundary condition. These values
are to be shifted by, say, $A_{\pm h}$ respectively,
so that on the horizon one obtains the physical $R$ and $T$.
We first correct the reflection coefficient on the horizon as follows:
Let $A_{-h}$ be the value of $A_-$ on the horizon (see, eq. 34),
$$
A_{-h}= - {c \over 2} + {\sqrt{[2k(r_+) - c^{2}]} \over 2}.
$$
It is appropriate to use ${\cal A}_-=A_- - A_{-h}$  rather than $A_-$ since
${\cal A}_-$ vanishes at $r=r_+$.

Incorporating these conditions, the solution (31) becomes,
$$
Z_{+} = \frac{{\cal A}_+} {\surd q} exp (i u) +
\frac {{\cal A}_-} {\surd q} exp (- i u) .
\eqno{(35)}
$$
with the usual normalization  condition
$$
{\cal A}_+^2+{\cal A}_-^2 = q .
\eqno{(36)}
$$
where, ${\cal A}_+=A_+-A_{+h}$.

Determination of $A_{+h}$ is done by enforcing $R$ obtained from eq. (37a),
which is shown below,
is the same as that obtained by the actual WKB method.
The $q$ is used to compute the transmission coefficient $T$ from eq. (36).
In this way, normalization of $R+T=1$ is assured.

The normalization factor $q \rightarrow k$ as $\hat{r}_{*} \rightarrow \infty$
and the condition $\frac{1}{q}\frac{dq}{d\hat{r}_{*}} << q$ is found
to be satisfied whenever $\frac{1}{k}\frac{dk}{d\hat{r}_{*}} << k$ is satisfied.
This is the essence of our modification of the WKB. In a true WKB, $A_\pm$ are
constants and the normalization is with respect to a (almost) constant $k$.
However, we are using it as if the WKB is instantaneously valid everywhere.
Our method may therefore be called `Instantaneous' WKB approximation
or IWKB for short. Using the new notations, the instantaneous values (i.e., 
local values)
of the reflection and transmission coefficients are given by (see, eq. 35),
$$
R= \frac {{\cal A}_-^2} { q}
\eqno{(37a)}
$$
$$
T= \frac {{\cal A}_+^2} { q} .
\eqno{(37b)}
$$
Whatever may be the value of the physical parameters,
$\frac{1}{k}\frac{dk}{d\hat{r}_{*}} << k$ is satisfied in whole range of
$\hat{r}_*$ for region I.

The variation of reflection and transmission coefficients would be
well understood if we imagine the potential barrier consists of a
large number of steps. From simple quantum mechanics, in between each two steps,  
we can calculate the reflection and transmission coefficients [9]. Clearly
these reflection and transmission coefficients at different junctions
will be different. This is discussed in detail below.
To be concrete, we choose one set of parameters from Region I. 
Here, the total energy of 
the incoming particle is greater than the potential barrier height for all 
values of $\hat{r}_*$.  We use,	Kerr parameter, $a = 0.5$; mass of the 
black hole, $ M = 1$; Mass of the particle, $ m_p = 0.8$; orbital 
angular momentum quantum number, $l = {1 \over 2}$; azimuthal 
quantum number, $m = - {1 \over 2}$; 
frequency of the incoming wave, $ \sigma = 0.8$. The derived parameters 
are, $ r_{+} = M + \surd(M^{2} - a^{2}) \cong 1.86603$; $\sigma_c = 1$; 
$\sigma_s = 0.066987$; $\alpha^{2} = - 0.0625$. For these parameters, 
the eigenvalue is $\lh = 0.92$ [7].

Here it is clear that $\sigma$ is in between $\sigma_c$ and $\sigma_s$
and $\alpha^{2} < 0$, $r_{+} > \vert \alpha \vert$. 
So we are in a strictly non super-radiant regime since 
here, $\sigma > \sigma_s$ [4].

>From eq. (24) we observe that there are two wave equations for
two potentials $V_{+}$ and $V_{-}$. The nature of the potentials
is shown in Fig. 2a. It is clear from the Fig. 2a that the potentials
$V_\pm$ are well behaved. They are monotonically decreasing as the
particle approaches the black hole, and the total energy chosen
in this case ($\sigma^2$) is always higher than $V_\pm$.
For concreteness, we study the equation with potential $V_{+}$.
A similar procedure (IWKB method) as explained above can be adopted 
using the potential $V_{-}$ to compute $Z_-$ and its form would be
$$
Z_{-} = \frac{A'_{+}-A'_{+h}} {\surd q'} exp (i u') - \frac {A'_{-}-A'_{-h}} {\surd q'} exp (- i u') .
\eqno{(35')}
$$
Note the occurance of the negative sign in front of the reflected wave. This is to
satisfy the asymptotic property of the wave functions. 

In Fig. 2b, we show the nature of $V_{+}$ (solid curve), $k$ (dashed curve) 
and $E (= \sigma^2)$ (short-dashed curve). In the 
solutions (eq. $35$ and $35'$) the first 
term corresponds to the incident wave and the second term corresponds to the 
reflected wave.  

In Fig. 2c, the variation of reflection and transmission coefficients are shown. 
It is seen that as matter comes close to the black hole,
the barrier height goes down. As a result, the penetration probability
increases, causing the rise of the transmission coefficient. 

Local values of the reflection and transmission co-efficients could also be 
calculated using the well known quantum mechanical approach. First one has to 
replace the potentials (as shown in Fig. 2a) by a collection of step functions 
as shown in Fig. 3a. The standard junction conditions of the type,
$$
Z_{{+},n}=Z_{{+},n+1}
\eqno{(38a)}
$$
where,
$$
Z_{+,n}  = A_{n} exp [ik_{n} {\hat r}_{*, n}] + B_{n} exp [-ik_n {\hat r}_{*, n}]
$$
and
$$
\frac{dZ_{+}}{d{\hat r}_*}|_n = \frac{dZ_{+}}{d{\hat r}_*}|_{n+1}
\eqno{(38b)}
$$
where,
$$
\frac{dZ_{+}}{d{\hat r}_*}|_n  = ik_n A_n exp (ik_n {\hat r}_{*,n}) - 
ik_n B_n exp (-ik_n{\hat r}_{*,n})
$$
at each of the $n$ steps were used to connect solutions at successive steps.
>From the simple quantum mechanical calculation we obtain the reflection
and transmission coefficients at the each junctions. 
Clearly at different junctions i.e., at different radii
this reflection and transmission coefficients will be different.
As before, we use the inner boundary condition, to be $R \rightarrow 0$
at ${\hat r}_* \rightarrow -\infty $. In reality we use as many steps as possible
to follow accurately the shape of the potential. Smaller
step sizes were used whenever $k$ varies faster. Fig. 3b shows the comparison of the
instantaneous reflection coefficients in both the methods. The agreement 
shows that the WKB can be used at each point quite successfully.

It is to be noted, that, strictly speaking, the terms
`reflection' and `transmission' coefficients are traditionally
defined with respect to the asymptotic values. The spatial 
dependence that we show are just the dependence of the instantaneous values.
This is consistent with the spirit of IWKB approximation that we are using.

The radial wave functions $R_{+ {1 \over 2}}$ and $R_{- {1 \over 2}}$ which are 
of spin up and spin down particles respectively of the original Dirac equation are given below,
$$
Re(R_{\frac{1}{2}} \Delta^{\frac{1}{2}}) = \frac{a_+ {\rm cos}(u - \theta) +
a_- {\rm cos}(u + \theta)}{2\sqrt{k}} + \frac{a'_+ {\rm cos}(u' - \theta) - a'_- {\rm cos}
(u' + \theta)}{2\sqrt{k'}}
\eqno{(39a)}
$$
$$
Im(R_{\frac{1}{2}} \Delta^{\frac{1}{2}}) = \frac{a_+ {\rm sin}(u - \theta) -
a_- {\rm sin}(u + \theta)}{2\sqrt{k}} + \frac{a'_+ {\rm sin}(u' - \theta) 
+ a'_- {\rm sin}(u' + \theta)}{2\sqrt{k'}}
\eqno{(39b)}
$$
$$
Re(R_{-\frac{1}{2}}) = \frac{a_+ {\rm cos}(u + \theta) +
a_- {\rm cos}(u - \theta)}{2\sqrt{k}} - \frac{a'_+ {\rm cos}(u' + \theta) 
- a'_- {\rm cos}(u' - \theta)} {2\sqrt{k'}}
\eqno{(39c)}
$$
$$
Im(R_{-\frac{1}{2}}) = \frac{a_+ {\rm sin}(u + \theta) -
a_- {\rm sin}(u - \theta)}{2\sqrt{k}} - \frac{a'_+ {\rm sin}(u' + \theta) 
+ a'_- {\rm sin}(u' - \theta)} {2\sqrt{k'}}
\eqno{(39d)}
$$
Here, $a_+=(A_+-A_{+h})/\sqrt(q/k)$ and $a_-=(A_--A_{-h})/\sqrt(q/k)$. 
$\frac{a'_+}{\sqrt{k'}}$ and $\frac{a'_-}{\sqrt{k'}}$ are the transmitted 
and reflected amplitudes respectively for the wave of corresponding potential $V_-$. 

In Fig. 4(a-d) we show the nature of these wavefunctions. The eikonals used
in plotting these functions (see, eq. 39[a-d]) have been calculated by approximating
$V_\pm$  in terms of a polynomial and using the 
definition $u ({\hat r}_*) = \int \sqrt(\sigma^2-V_\pm) d {\hat r}_* $.
This was done since $V_\pm$ is not directly integrable. Note that the amplitude as
well as wavelength remain constants in regions where $k$ is also constant. 

\subsection*{Solutions of Region II}

Here we study the solution of a region  where for any set of physical
parameters, the total energy of the incoming particle is less than the maximum
height of the potential barrier. So the WKB approximation (more precisely,
our IWKB approximation) is not valid in the whole 
range of $\hat{r}_*$. In the regions where the 
WKB is not valid, the solutions will be the
linear combination of Airy functions because the 
potential is a linear function  of ${\hat{r}}_*$ in those 
intervals.  At the junctions one has to match the solutions 
including Airy functions with the solution obtained by WKB method. 

In the region where the WKB approximation is valid, 
local values of reflection and transmission
coefficients and the wave function can be calculated easily by following 
the same method described in previous sub-section (solutions of region I)
and the solution will be same as equation ($35, 35'$). 
In other regions, the equation reduces to
$$
{{d^{2}Z_{+}} \over {d{\hat{r}}_{*}^{2}}} - x Z_{+} = 0 
\eqno{(40)}
$$
where $x = \beta^{1 \over 3} ({\hat{r}}_{*} - p)$, $\beta$ is 
chosen to be positive and $p$ is the critical
point where the total energy and potential energy are matching.

Let $Z_{+}(x) = x^{1 \over 2} Y(x)$
and considering region $x > 0$ the equation (40) reduces to
$$
x^{2} {{d^{2}Y} \over {dx^{2}}} + x {{dY} \over {dx}} - \left(x^{3} + 
{1 \over 4}\right) Y(x) = 0 .
\eqno{(41)}
$$

By making another transformation
$$
\xi = {2 \over 3} x^{3 \over 2}
\eqno{(42)}
$$
we obtain
$$
\xi^{2} {{d^{2}Y} \over {d\xi^{2}}} + \xi {{dY} \over {d\xi}} - 
\left(\xi^{2} + {1\over 9}\right) Y(\xi) = 0,
\eqno{(43)}
$$
this is the modified Bessel equation.
The solution of this equation is $I_{+ {1 \over 3}}(\xi)$ and $I_{- {1 \over 3}}(\xi)$.
So the solution of eq. (40) will be 
$$
Z_{+}(x) = x^{1 \over 2} [C_{1} I_{+ {1 \over 3}}(\xi) + C_{2} I_{- {1 \over 3}}(\xi)].
\eqno{(44)}
$$

When $x<0$ the corresponding equation is,
$$
\xi^{2} {{d^{2}Y} \over {d\xi^{2}}} + \xi {{dY} \over {d\xi}}
+ \left(\xi^{2} - {1 \over 9}\right) Y(\xi) = 0,
\eqno{(45)}
$$
which is the Bessel equation. The corresponding solution is
$$
Z_{+}(x) = | x |^{1 \over 2} [D_{1} J_{+ {1 \over 3}}(\xi) + D_{2} J_{- {1 \over 3}}(\xi)],
\eqno{(46)}
$$
where $J_{\pm\frac{1}{3}}$ and $I_{\pm}\frac{1}{3}$ 
are the Bessel functions and the modified Bessel 
functions of order $\frac{1}{3}$ respectively.

The Airy functions are defined as
$$
Ai(x) = {1 \over 3} x^{1 \over 2} [I_{- {1 \over 3}}(\xi) - I_{+ {1 \over
3}}(\xi)],  \hskip1cm x > 0 ,
\eqno{(47)}
$$

$$
Ai(x) = {1 \over 3} | x |^{1 \over 2} [J_{- {1 \over 3}}(\xi) + J_{+ {1 \over
3}}(\xi)], \hskip1cm x < 0 ,
\eqno{(48)}
$$

$$Bi(x) = {1 \over \surd{3}} x^{1 \over 2} [I_{- {1 \over 3}}(\xi) + I_{+ {1
\over 3}}(\xi)], \hskip1cm x > 0 ,
\eqno{(49)}
$$

$$
Bi(x) = {1 \over \surd{3}} | x |^{1 \over 2} [J_{- {1 \over 3}}(\xi) - J_{+ {1
\over 3}}(\xi)], \hskip1cm  x < 0 .
\eqno{(50)}
$$

In terms of Airy functions, the solutions (44) and (46) can be written as 
$$
Z_{+} = {3 \over 2} (C_{2} - C_{1}) Ai(x) + {\surd{3} \over 2} (C_{2} + C_{1}) Bi(x) 
\hskip0.5cm {\rm for}\hskip0.2cm x > 0,
\eqno{(51)}
$$
$$Z_{+} = {3 \over 2} (D_{2} + D_{1}) Ai(x) + {\surd{3} \over 2} (D_{2} - D_{1}) Bi(x) 
\hskip0.5cm {\rm for}\hskip0.2cm x < 0.
\eqno{(52)}
$$
By matching boundary conditions it is easy to show that the solution corresponding 
$x > 0$ and that corresponding $x < 0$ are continuous 
when $C_1 = - D_1$ and $C_2 = D_2$. 

As an example of solutions from this region, we choose: $a= 0.95$, $ M = 1$, $ m_p = 0.17$, 
$l =  {1 \over 2}$, $m = - {1 \over 2}$, and $ \sigma = 0.21$. 
The black hole horizon is at $ r_+ = M + \surd{(M^2 - a^2)} \cong 1.31225$,
$\sigma_c = 0.526316$, $\sigma_s = 0.180987$, $\alpha^{2} = - 1.356$ and $\lh = 0.93$ [7].
It is clear that the values of $\sigma_{c}$, $\sigma_{s}$ and $\alpha^{2}$
indicate the region is non super-radiant. 
In Fig. 5a, we show the nature of $V_+$ and $V_-$, however, while solving, we use
the equation containing $V_{+}$ (eq. 24). Unlike the case in the 
previous sub-section, here $\sigma^2$ is no longer
greater than $V_\pm$ at all radii. As  a result, 
$k^2$ may attain negative values in some region. 
In Fig. 5b, nature of $V_{+}$ (solid curve), parameter $k$ (dashed curve) and 
energy $E$ (short-dashed curve) are shown. 
Here, WKB approximation can be applied  in regions other than
${{\hat{r}}_{*}}  \sim  - 4$ to $- 1$ and $1$ to $7$ where $k$ is close to zero
and the condition ${1 \over k} {dk \over d{{\hat{r}}_{*}}} < < k$ is not satisfied.

In  the region  ${\hat{r}}_{*}=7$ to $1$ around the turning point ${\hat{r}}_{*} = 4.45475$
the solutions turns out to be [10],
$$
Z_{+} = 1.087526 Ai(x) + 0.788968 Bi(x).
\eqno{(53)}
$$
Similarly, the solution from ${\hat{r}}_{*} \sim - 1$ to $- 4$ i.e. 
around the turning point 
${\hat{r}}_{*} = - 2.8053$ turns out to be [10],
$$
Z_{+} = - [1.328096 Ai(x) + 0.774426 Bi(x)].
\eqno{(54)}
$$
It is to be noted that in the region ${\hat{r}}_{*} \sim 1$ to $-1$, even though the 
potential energy dominates over the total energy, WKB approximation method is 
still valid. Here the solution will take the form 
$\frac {exp(- u)} {\surd k}$ and $\frac {exp(+ u)} {\surd k}$. 
Asymptotic values of the instantaneous reflection and the transmission coefficients 
(which are traditionally known as the `reflection' and `transmission' coefficients) are 
obtained from the WKB approximation. This yields the integral constant $c$ 
as in previous case.
>From eq. 37(a-b) local reflection and transmission coefficients 
are calculated, behaviours of which are shown in Fig. 5c. The constants $A_{-h}$ 
and $A_{+h}$ are calculated as before. Note the decaying nature
of the reflection coefficient inside the potential barrier.

\subsection*{Solutions in the Super-Radiant Region}

In this region, the potential diverges at $r=|\alpha|$. Here, the barrier
height goes up to infinity and then the potential changes sign so that
its nature changes from repulsive to attractive and vice versa. This
is because $\sigma < \sigma_s$ (which is the condition
for super-radiance) and $r_+<|\alpha|$ [see equations (11) and (13)]. 
Unlike the previous two cases, the relation between 
$r$ and ${\hat r}_*$ is not single valued.
Here, at both $r = r_+$ and $r = \infty$, the value of ${\hat r}_* = \infty$. 
With the decrement of $r$, ${\hat r}_*$ is decreased initially up to 
$r = | \alpha |$. Subsequently, ${\hat r}_*$ 
starts to rise and at the black hole horizon it diverges.
Obviously, in this case particles hit the barrier and we can solve the
equation following the same methods as explained in the previous cases, i.e.,
eqs. (35) and (35$'$) for the region where the WKB method
is valid and eqs. (51) and (52) where the WKB method is not valid. 

For illustrative example, 
here, we choose: $a = 0.95 $, $M = 1 $,  $m_p = 0.105 $,
$l =  {1 \over 2}$, $m = - {1 \over 2}$, and  $\sigma = 0.105. $
The  black hole horizon is located at $r_+ = M + \surd(M^2 - a^2) \cong 1.31225 $, 
and $\sigma_c = 0.526316 $, $\sigma_s = 0.180987 $, $\alpha^2 = - 3.62$ 
and $\lh = 0.97$ [7]. Chandrasekhar showed [4] that for integral spin
particles this region exhibits super-radiance and conjectured that for half-integral
spins the super-radiance may be absent. We investigate here if this conjecture is
valid.

The behaviour of potentials $V_+$ and $V_-$ are shown in Fig. 6a. 
It is clear that at $r = | \alpha |$ the potential diverges and
the nature of the potential is changed from repulsive to attractive (for $V_-$)
and vice versa (for $V_+$). Here, we will 
treat the equation with $V_-$ as the potential (it is equally easy to do the 
problem with $V_+$). We first divide our computations into two parts.
In the repulsive part of the potential (i.e., when $V_->0$), particles come
from infinity and most of them reflect back from the infinitely high barrier. 
In the attractive part of the potential (i.e., when $V_+<0$),  
particle radiates outwards in the ${\hat r}_*$ co-ordinate 
(actually, particle goes towards the horizon but due to multivalueness 
of the radial co-ordinate ${\hat r}_{*}$ (with respect to $r$) the horizon 
is mapped to infinity). 

In Fig. 6b, nature of $V_{-}$, $k$ and $E$ are shown.
The WKB approximation (more precisely IWKB approximation) 
method is valid from infinity to ${\hat r}_* \sim  40$ 
since, otherwise, $\frac{1}{k} \frac{dk}{d\hat{r}_*} <<k$ is not 
satisfied. In those other regions one has to apply a different method (which was
also explained in last sub-section) to find solutions. 
The local values of the reflection and transmission coefficients 
and the wave function of the particle are calculated as in the 
previous cases. Since the matter which tunnels through the infinitely high
barrier face infinitely strong attractive field,
the possibility of extraction of energy would be zero. 
In Fig. 6c, the variations of local transmission and reflection coefficients are shown.
The net transmission of the wave through the horizon is non-negative all along 
and therefore super-radiation is absent, although $\sigma$ is less than $\sigma_{s}$. 
We believe that the non-existence of super-radiation is due to  
($r - | \alpha |)^{-3}$ variation of the potential near the singular point. 
Because of the existence of attractive field, the
extraction of energy is difficult, so the net transmission of the wave
through the horizon from $\infty$ is always positive. This argument is
valid for any set of parameters where $\sigma \leq \sigma_{s}$. 

\section*{IV. CONCLUSION}

In this paper, we studied scattering of massive, spin-half particles from a 
Kerr black hole, particularly the nature of the radial wave functions and the
reflection and transmission coefficients. Our main motivation was to give a 
general analytical expression of the solution which can be useful for further study.
We showed few illustrative cases as examples. 
We verified that these analytical solutions
were indeed correct by explicitly solving the same set of equations 
using quantum mechanical step-potential approach as described in Section III. We classified
the entire parameter space in terms of the physical and unphysical regions
and the physical region was further classified into two regions,
depending on whether the particle `hits' the potential barrier or not.
Again, the region where particle hits on the barrier, is divided into two
parts, one is super-radiant region and other is non-super-radiant region. 
We chose one illustrative example in each of the regions. We emphasize that
the most `interesting' region to study would be close to $m_p \sim \sigma$.
However we pointed out (Fig. 1) that for $m_p \leq 0.35$, the WKB solutions
cannot be trusted, and other methods (such as those using Airy functions)
must be employed.

We used the well known WKB approximation method as well as the step-potential method
of quantum mechanics to obtain the spatial dependence 
of the coefficients of the wave function.
This in turn, allowed us to determine the reflection and transmission
coefficients and the nature of wave functions.
The usual WKB method with constant coefficients and (almost) constant wave number
$k$ is successfully applied even when the coefficients and  wave number are not constant
everywhere. Solution from this `instantaneous' WKB (or IWKB) method
agrees fully with that obtained from a purely quantum mechanical method
where the potential is replaced by a collection of steps.
Our method of obtaining solutions should be
valid for any black hole geometry which are asymptotically flat so that
radial waves could be used at a large distance.
This way we ensure that the analytical solution is close to the exact
solution. In Region II, in some regions, the WKB method cannot be applied and
hence Airy function approach or our step-potential approach could be used.

In the literature, reflection and transmission coefficients are defined
at a single point. These definitions  are meaningful only if the
potential varies in a small region while studies are made from a large distance
from it. In the present case, the potential changes over a large distance
and we are studying in these regions  as well.
Although we used the words `reflection' and `transmission' coefficients,
in this paper very loosely, our definitions are very rigorous and well
defined. These quantities are simply the instantaneous values and in our belief
more physical. The problem at hand is very
similar to the problem of reflection and transmission of acoustic
waves from a strucked string of non-constant density where reflection and transmission
occurs at each point.

Among other things, we verify Chandrasekhar's conjecture [4] 
based on the asymptotic solution,
that for spin-$1 \over 2$ particle the phenomenon of super-radiance is absent.
We believe that this is due to the very way the potential develops
the singularity at $r = | \alpha |$.  Here $V_-({\hat r}_*)
\propto (r - | \alpha |)^{-3}$, which results in an
attractive potential in some region very close to the black hole. In contrast,
$V_-({\hat r}_*) \propto (r - | \alpha |)^{-4}$ when electromagnetic
and gravitational waves are scattered off the black hole [4] does not
create an attractive part in the potential and possibly exhibit the 
phenomenon of super-radiance. It is noted that all the cases where
potential diverge at $r=\alpha$ (i.e., so called super-radiation cases)
arise for $\sigma \leq \sigma_s$ with the negative values of 
azimuthal quantum number (here, $m=-1/2$) and the positive Kerr parameter, $a$.
For positive values of $m$ and positive values of $a$, potential does not
diverge at any point for all values of $\sigma$. If we change 
the spin orientation of the black hole (negative values of $a$) and take
positive $m$ again divergence of the potential will arise. Thus, it seems
that the  cases with opposite sign of $a$ and $m$ are physically more interesting.  

It is seen that for different physical parameters the solutions are different.
The waves scattered off are distinctly different in different parameter regions.
In a way, therefore, black holes can act as a mass spectrograph! Another interesting application
of our method would be to study interactions of Hawking radiations in regions 
just outside the horizon.


\section*{FIGURE CAPTIONS}

\begin{figure}

\noindent Fig. 1: Contours of constant $w_{max}$=
max$(\frac{1}{k^2}\frac{dk}{d{\hat{r}}_*})$
are shown to indicate that generally $w<<1$ and therefore the WKB approximation
is valid in most of the physical region. Labels indicate values of $w_{max}$.
\end{figure}

\begin{figure}
\noindent Fig. 2: Behaviour of (a) $V_+$ (solid curve) and $V_-$ (dashed curve),
(b) $V_+$ (solid curve), $k$ (dashed curve), total energy $E$ (short-dashed curve),
(c) local transmission ($T$, solid curve) and reflection
($R$, dashed curve) coefficients as functions of $\hat{r}_*$.  The
parameters are $a=0.5$, $M=1$, $m_p=0.8$, $l=\frac{1}{2}$, $m=-\frac{1}{2}$, $\sigma=0.8$.
\end{figure}

\begin{figure}
\noindent Fig. 3a: Steps (solid) approximating a potential (dotted), thus 
reducing the problem to that of a quantum mechanics. 
The parameters are same as in Fig. 2.
\end{figure}

\begin{figure}
\noindent Fig. 3b: Comparison of variation of instantaneous reflection
coefficient $R$ and transmission coefficient $T$ with the 
radial coordinate ${\hat r}_*$ using analytical WKB method (solid) 
and step-potential method (dotted). The parameters are same as Fig. 2.
\end{figure}

\begin{figure}
\noindent Fig. 4: Nature of real and imaginary parts
of radial wave functions for Case 1. 
\end{figure}

\begin{figure}
\noindent Fig. 5: Plots are same as in Fig. 2.
The parameters are $a=0.95$, $M=1$, $m_p=0.17$, $l=\frac{1}{2}$, $m=-\frac{1}{2}$,
$\sigma=0.21$.
\end{figure}

\begin{figure}
\noindent Fig. 6:  Behaviour of (a) $V_+$ (solid curve) and $V_-$ (dashed curve),
(b) $V_-$ (solid curve), $k$ for region where potential is positive ( $k_{rep}$,
dashed curve), $k$ for region where potential is negative ( $k_{att}$,
short-dashed curve), total energy $E$ (dotted curve),
(c) local transmission ($T$, solid curve) and reflection ($R$, dashed curve) coefficients
as functions of $\hat{r}_*$. 
The parameters are $a=0.95$, $M=1$, $m_p=0.105$, $l=\frac{1}{2}$, $m=-\frac{1}{2}$,
$\sigma=0.105$.
\end{figure}

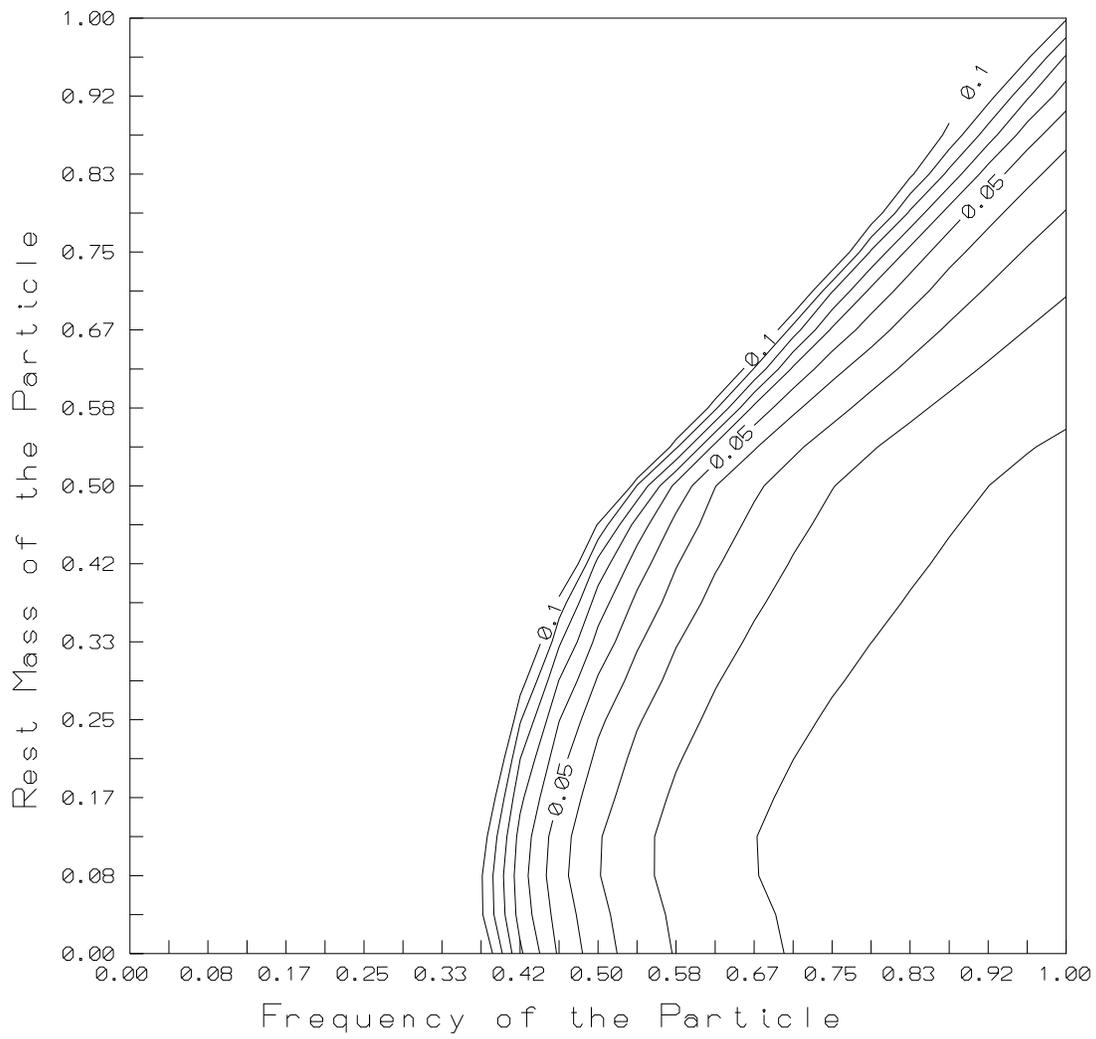
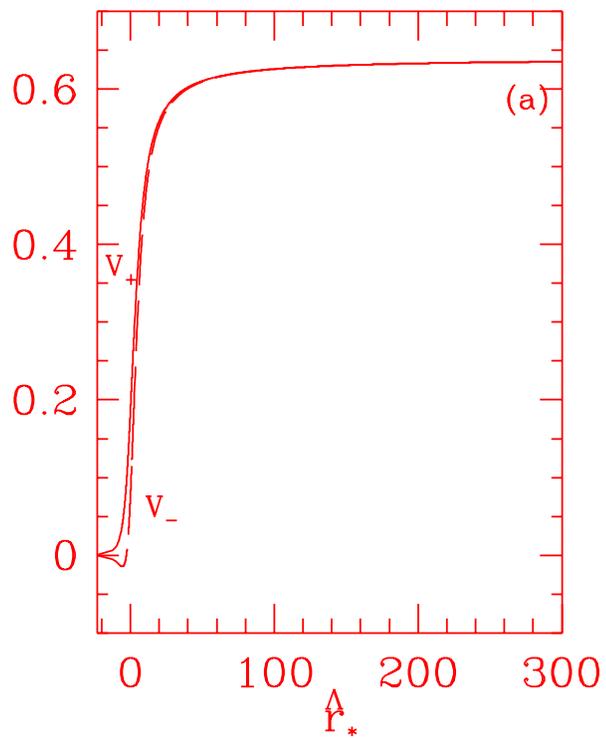
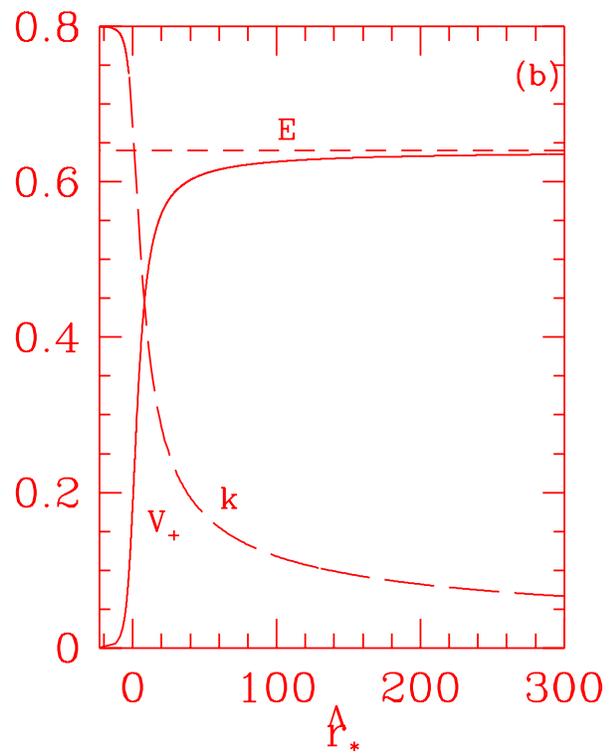
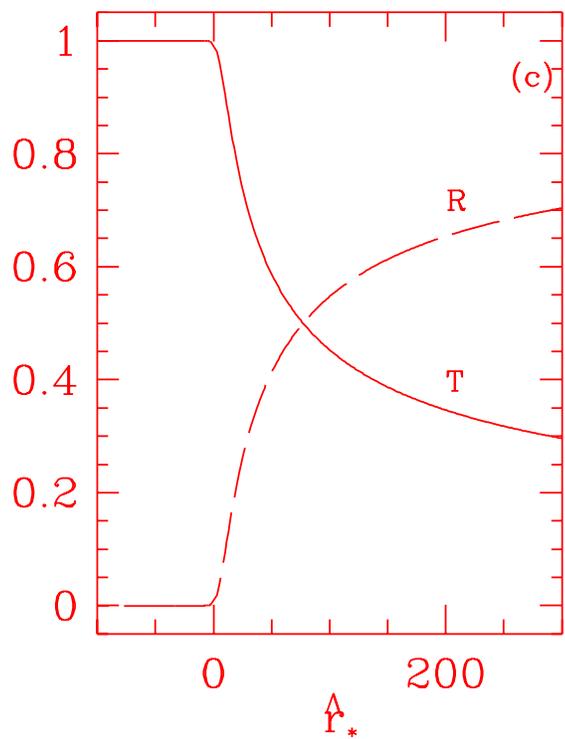
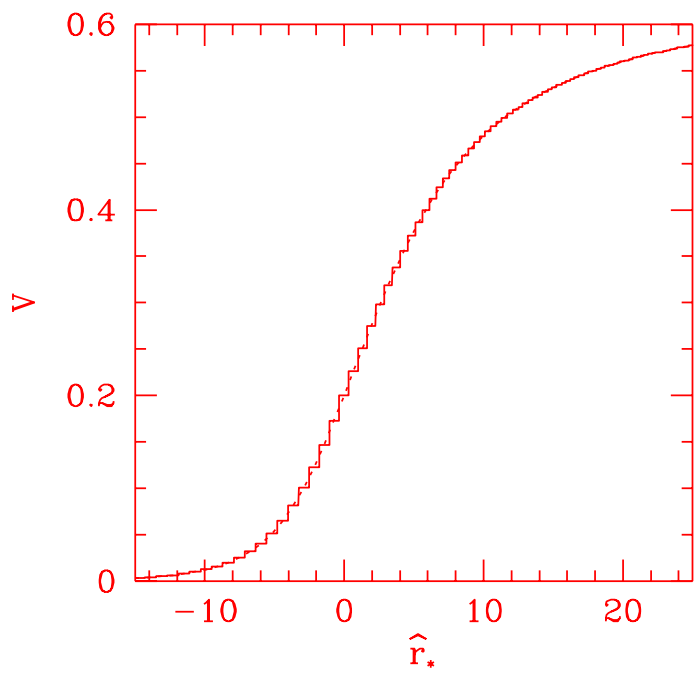
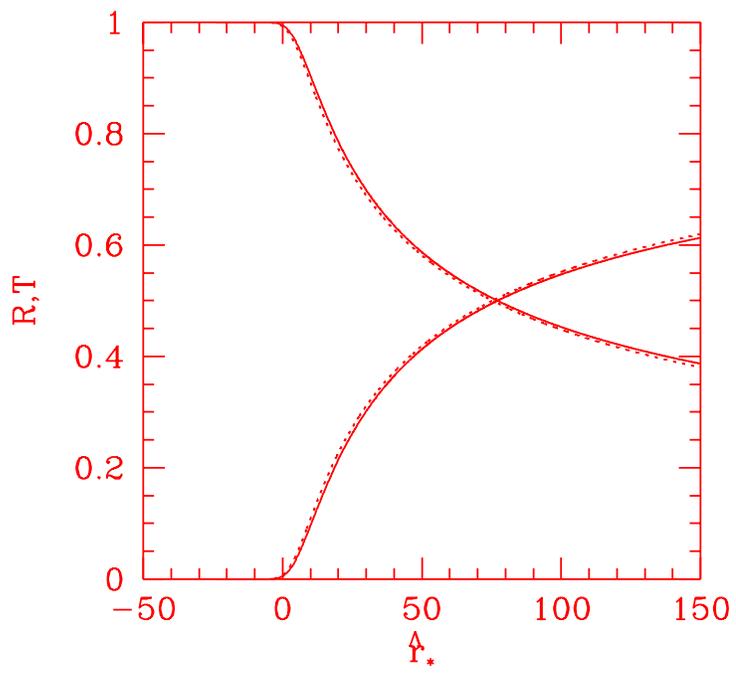
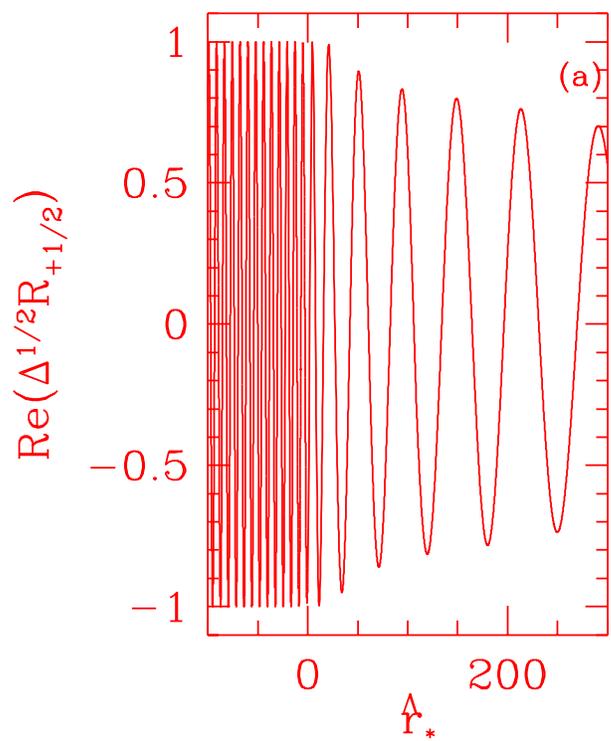
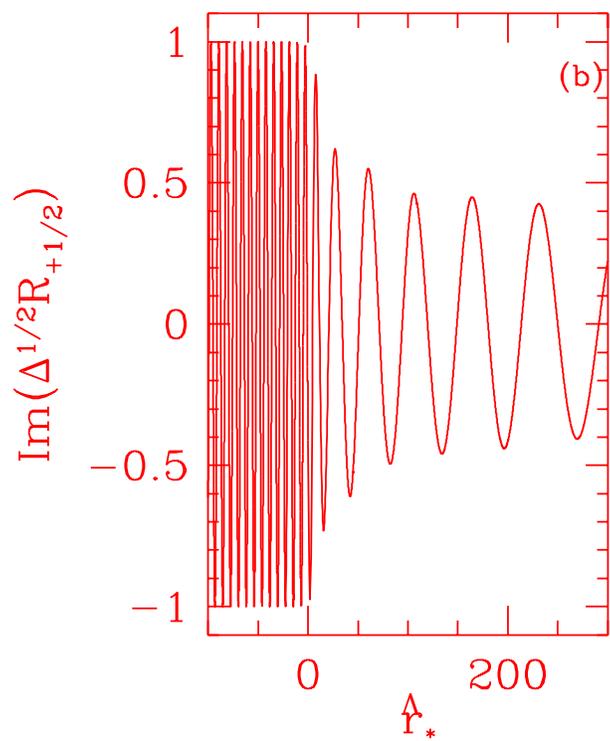
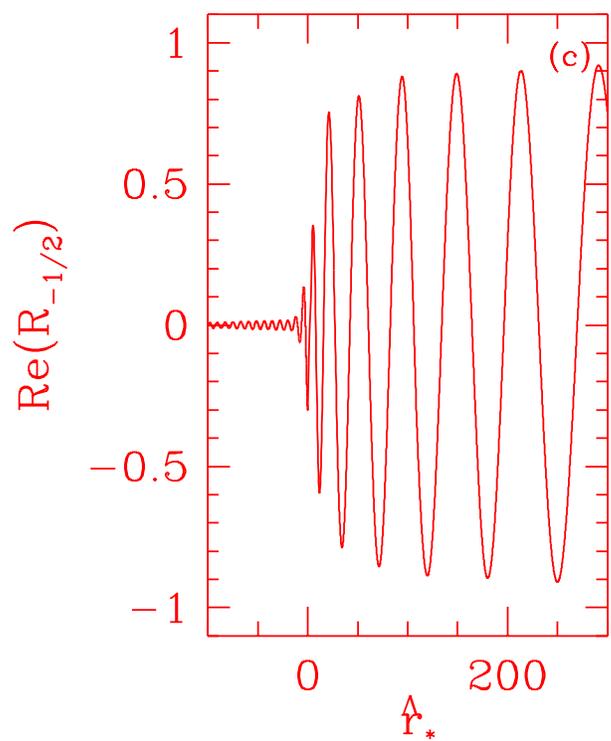
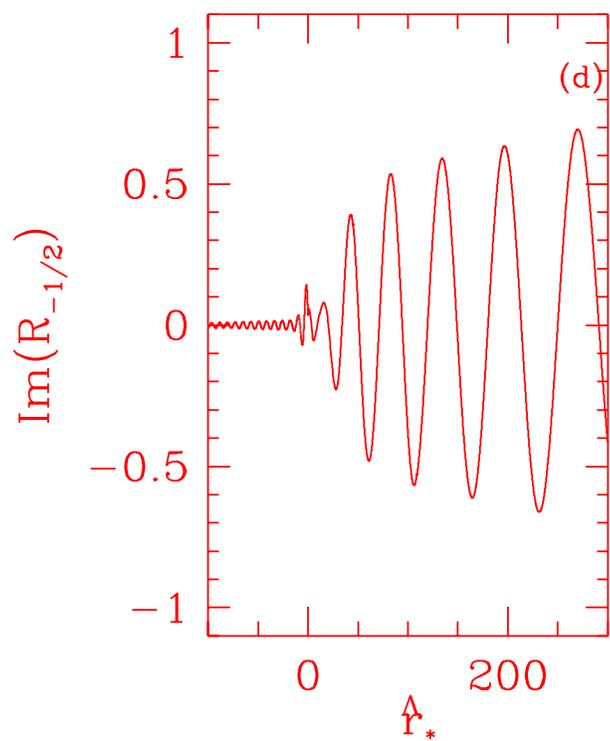
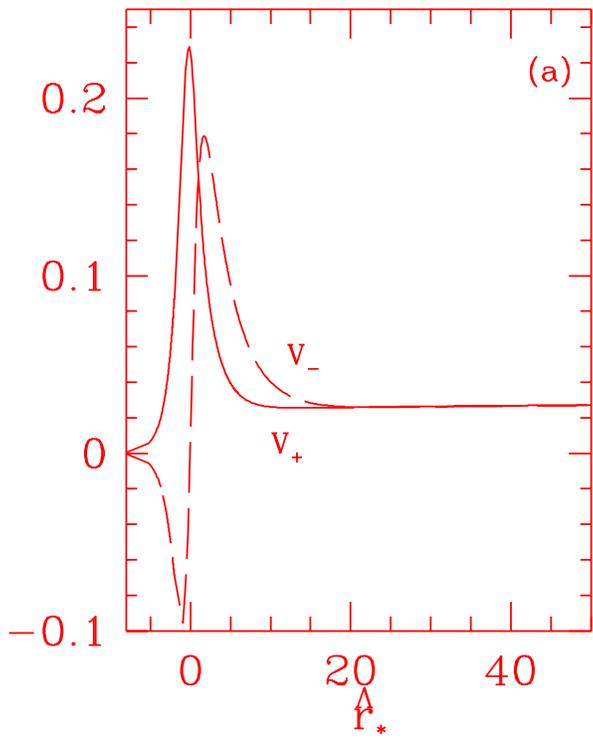
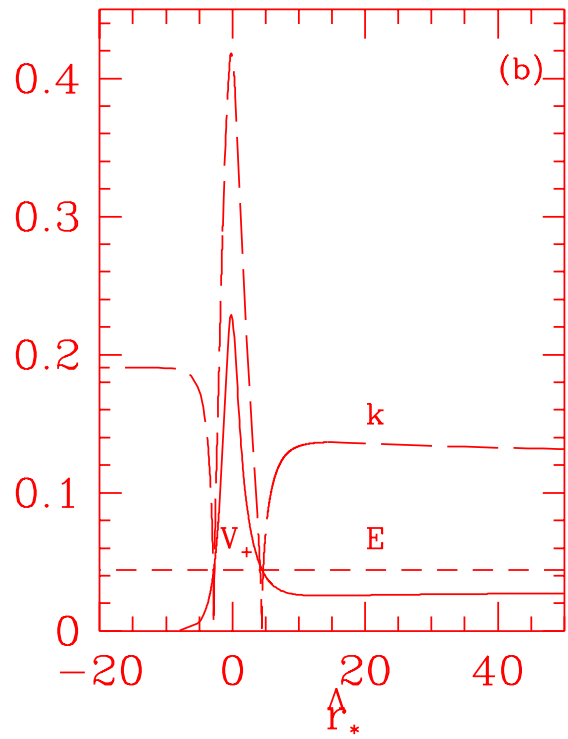
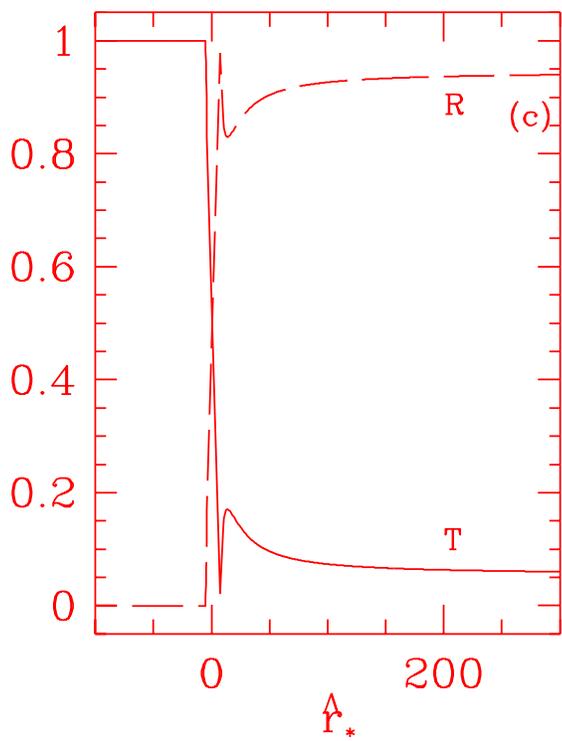
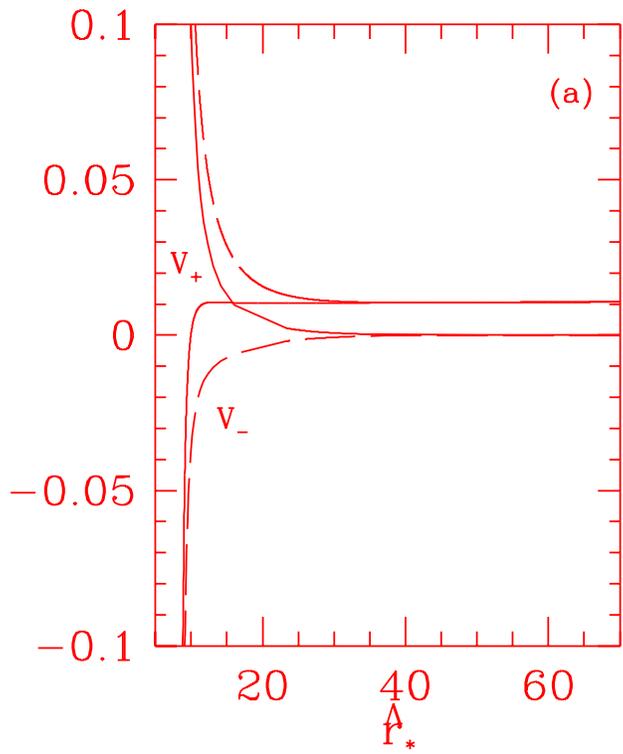
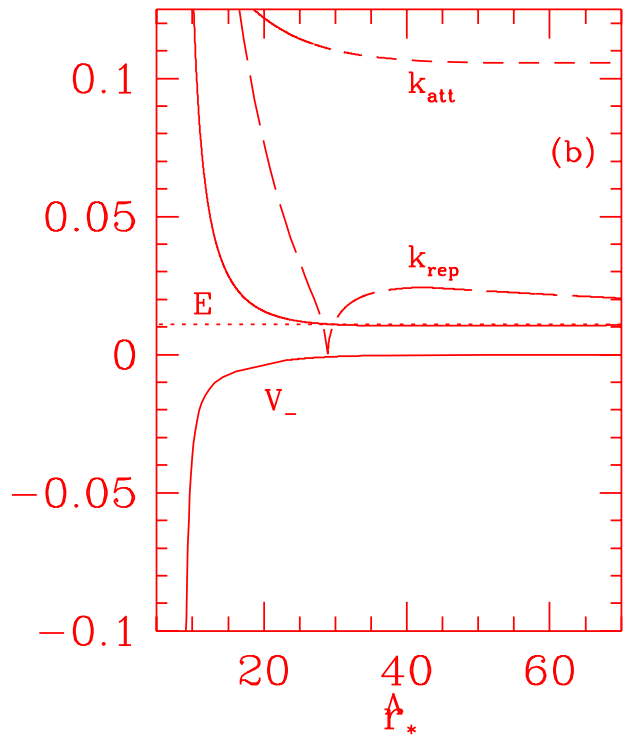
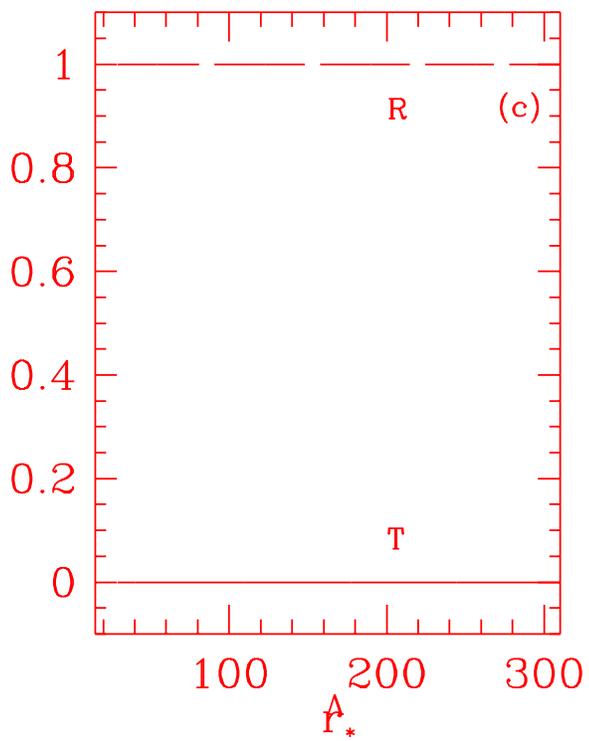
\end{document}